\RequirePackage{lineno}

\documentclass[twocolumn,showpacs,aps,prl,superscriptaddress]{revtex4}

\usepackage{graphicx}
\usepackage{dcolumn}
\usepackage{bm}
\usepackage{slashed}
\usepackage{subfig}

\RequirePackage{xspace}
\usepackage{relsize}

\def\Kbar  {\kern 0.2em\overline{\kern -0.2em K}{}\xspace}

\def\Bbar    {\kern 0.18em\overline{\kern -0.18em B}{}\xspace}

\def\Qbar    {\kern 0.08em\overline{\kern -0.08em Q}{}\xspace}

\def\invfb   {\ensuremath{\mbox{\,fb}^{-1}}\xspace}
\newcommand{\mev}{\ensuremath{\mathrm{\,Me\kern -0.1em V}}\xspace}
\newcommand{\mevc}{\ensuremath{{\mathrm{\,Me\kern -0.1em V\!/}c}}\xspace}
\newcommand{\mevcc}{\ensuremath{{\mathrm{\,Me\kern -0.1em V\!/}c^2}}\xspace}
\newcommand{\gev}{\ensuremath{\mathrm{\,Ge\kern -0.1em V}}\xspace}
\newcommand{\gevc}{\ensuremath{{\mathrm{\,Ge\kern -0.1em V\!/}c}}\xspace}
\newcommand{\gevcnospace}{\ensuremath{{\mathrm{\,Ge\kern -0.1em V\!/}c}}}
\newcommand{\gevcc}{\ensuremath{{\mathrm{\,Ge\kern -0.1em V\!/}c^2}}\xspace}

\def\missET {\mbox{\ensuremath{\not\!\! E_T}}}

\begin{document}
\bibliographystyle{apsrev}

\title{{Search for Production of Heavy Particles Decaying to
 Top Quarks  \\and Invisible
    Particles in $p\bar{p}$ collisions at $\sqrt{s}=1.96$ TeV}}

\affiliation{Institute of Physics, Academia Sinica, Taipei, Taiwan 11529, Republic of China} 
\affiliation{Argonne National Laboratory, Argonne, Illinois 60439, USA} 
\affiliation{University of Athens, 157 71 Athens, Greece} 
\affiliation{Institut de Fisica d'Altes Energies, ICREA, Universitat Autonoma de Barcelona, E-08193, Bellaterra (Barcelona), Spain} 
\affiliation{Baylor University, Waco, Texas 76798, USA} 
\affiliation{Istituto Nazionale di Fisica Nucleare Bologna, $^{aa}$University of Bologna, I-40127 Bologna, Italy} 
\affiliation{University of California, Davis, Davis, California 95616, USA} 
\affiliation{University of California, Irvine, Irvine, California 92697, USA} 
\affiliation{University of California, Los Angeles, Los Angeles, California 90024, USA} 
\affiliation{Instituto de Fisica de Cantabria, CSIC-University of Cantabria, 39005 Santander, Spain} 
\affiliation{Carnegie Mellon University, Pittsburgh, Pennsylvania 15213, USA} 
\affiliation{Enrico Fermi Institute, University of Chicago, Chicago, Illinois 60637, USA}
\affiliation{Comenius University, 842 48 Bratislava, Slovakia; Institute of Experimental Physics, 040 01 Kosice, Slovakia} 
\affiliation{Joint Institute for Nuclear Research, RU-141980 Dubna, Russia} 
\affiliation{Duke University, Durham, North Carolina 27708, USA} 
\affiliation{Fermi National Accelerator Laboratory, Batavia, Illinois 60510, USA} 
\affiliation{University of Florida, Gainesville, Florida 32611, USA} 
\affiliation{Laboratori Nazionali di Frascati, Istituto Nazionale di Fisica Nucleare, I-00044 Frascati, Italy} 
\affiliation{University of Geneva, CH-1211 Geneva 4, Switzerland} 
\affiliation{Glasgow University, Glasgow G12 8QQ, United Kingdom} 
\affiliation{Harvard University, Cambridge, Massachusetts 02138, USA} 
\affiliation{Division of High Energy Physics, Department of Physics, University of Helsinki and Helsinki Institute of Physics, FIN-00014, Helsinki, Finland} 
\affiliation{University of Illinois, Urbana, Illinois 61801, USA} 
\affiliation{The Johns Hopkins University, Baltimore, Maryland 21218, USA} 
\affiliation{Institut f\"{u}r Experimentelle Kernphysik, Karlsruhe Institute of Technology, D-76131 Karlsruhe, Germany} 
\affiliation{Center for High Energy Physics: Kyungpook National University, Daegu 702-701, Korea; Seoul National University, Seoul 151-742, Korea; Sungkyunkwan University, Suwon 440-746, Korea; Korea Institute of Science and Technology Information, Daejeon 305-806, Korea; Chonnam National University, Gwangju 500-757, Korea; Chonbuk National University, Jeonju 561-756, Korea} 
\affiliation{Ernest Orlando Lawrence Berkeley National Laboratory, Berkeley, California 94720, USA} 
\affiliation{University of Liverpool, Liverpool L69 7ZE, United Kingdom} 
\affiliation{University College London, London WC1E 6BT, United Kingdom} 
\affiliation{Centro de Investigaciones Energeticas Medioambientales y Tecnologicas, E-28040 Madrid, Spain} 
\affiliation{Massachusetts Institute of Technology, Cambridge, Massachusetts 02139, USA} 
\affiliation{Institute of Particle Physics: McGill University, Montr\'{e}al, Qu\'{e}bec, Canada H3A~2T8; Simon Fraser University, Burnaby, British Columbia, Canada V5A~1S6; University of Toronto, Toronto, Ontario, Canada M5S~1A7; and TRIUMF, Vancouver, British Columbia, Canada V6T~2A3} 
\affiliation{University of Michigan, Ann Arbor, Michigan 48109, USA} 
\affiliation{Michigan State University, East Lansing, Michigan 48824, USA}
\affiliation{Institution for Theoretical and Experimental Physics, ITEP, Moscow 117259, Russia}
\affiliation{University of New Mexico, Albuquerque, New Mexico 87131, USA} 
\affiliation{Northwestern University, Evanston, Illinois 60208, USA} 
\affiliation{The Ohio State University, Columbus, Ohio 43210, USA} 
\affiliation{Okayama University, Okayama 700-8530, Japan} 
\affiliation{Osaka City University, Osaka 588, Japan} 
\affiliation{University of Oxford, Oxford OX1 3RH, United Kingdom} 
\affiliation{Istituto Nazionale di Fisica Nucleare, Sezione di Padova-Trento, $^{bb}$University of Padova, I-35131 Padova, Italy} 
\affiliation{LPNHE, Universite Pierre et Marie Curie/IN2P3-CNRS, UMR7585, Paris, F-75252 France} 
\affiliation{University of Pennsylvania, Philadelphia, Pennsylvania 19104, USA}
\affiliation{Istituto Nazionale di Fisica Nucleare Pisa, $^{cc}$University of Pisa, $^{dd}$University of Siena and $^{ee}$Scuola Normale Superiore, I-56127 Pisa, Italy} 
\affiliation{University of Pittsburgh, Pittsburgh, Pennsylvania 15260, USA} 
\affiliation{Purdue University, West Lafayette, Indiana 47907, USA} 
\affiliation{University of Rochester, Rochester, New York 14627, USA} 
\affiliation{The Rockefeller University, New York, New York 10065, USA} 
\affiliation{Istituto Nazionale di Fisica Nucleare, Sezione di Roma 1, $^{ff}$Sapienza Universit\`{a} di Roma, I-00185 Roma, Italy} 

\affiliation{Rutgers University, Piscataway, New Jersey 08855, USA} 
\affiliation{Texas A\&M University, College Station, Texas 77843, USA} 
\affiliation{Istituto Nazionale di Fisica Nucleare Trieste/Udine, I-34100 Trieste, $^{gg}$University of Trieste/Udine, I-33100 Udine, Italy} 
\affiliation{University of Tsukuba, Tsukuba, Ibaraki 305, Japan} 
\affiliation{Tufts University, Medford, Massachusetts 02155, USA} 
\affiliation{University of Virginia, Charlottesville, VA  22906, USA}
\affiliation{Waseda University, Tokyo 169, Japan} 
\affiliation{Wayne State University, Detroit, Michigan 48201, USA} 
\affiliation{University of Wisconsin, Madison, Wisconsin 53706, USA} 
\affiliation{Yale University, New Haven, Connecticut 06520, USA} 
\author{T.~Aaltonen}
\affiliation{Division of High Energy Physics, Department of Physics, University of Helsinki and Helsinki Institute of Physics, FIN-00014, Helsinki, Finland}
\author{B.~\'{A}lvarez~Gonz\'{a}lez$^w$}
\affiliation{Instituto de Fisica de Cantabria, CSIC-University of Cantabria, 39005 Santander, Spain}
\author{S.~Amerio}
\affiliation{Istituto Nazionale di Fisica Nucleare, Sezione di Padova-Trento, $^{bb}$University of Padova, I-35131 Padova, Italy} 

\author{D.~Amidei}
\affiliation{University of Michigan, Ann Arbor, Michigan 48109, USA}
\author{A.~Anastassov}
\affiliation{Northwestern University, Evanston, Illinois 60208, USA}
\author{A.~Annovi}
\affiliation{Laboratori Nazionali di Frascati, Istituto Nazionale di Fisica Nucleare, I-00044 Frascati, Italy}
\author{J.~Antos}
\affiliation{Comenius University, 842 48 Bratislava, Slovakia; Institute of Experimental Physics, 040 01 Kosice, Slovakia}
\author{G.~Apollinari}
\affiliation{Fermi National Accelerator Laboratory, Batavia, Illinois 60510, USA}
\author{J.A.~Appel}
\affiliation{Fermi National Accelerator Laboratory, Batavia, Illinois 60510, USA}
\author{A.~Apresyan}
\affiliation{Purdue University, West Lafayette, Indiana 47907, USA}
% ...
\collaboration{CDF Collaboration\footnote{With visitors from $^a$University of Massachusetts Amherst, Amherst, Massachusetts 01003,
$^b$Istituto Nazionale di Fisica Nucleare, Sezione di Cagliari, 09042 Monserrato (Cagliari), Italy,
$^c$University of California Irvine, Irvine, CA  92697, 
$^d$University of California Santa Barbara, Santa Barbara, CA 93106
$^e$University of California Santa Cruz, Santa Cruz, CA  95064,
$^f$CERN,CH-1211 Geneva, Switzerland,
$^g$Cornell University, Ithaca, NY  14853, 
$^h$University of Cyprus, Nicosia CY-1678, Cyprus, 
$^i$University College Dublin, Dublin 4, Ireland,
$^j$University of Fukui, Fukui City, Fukui Prefecture, Japan 910-0017,
$^k$Universidad Iberoamericana, Mexico D.F., Mexico,
$^l$Iowa State University, Ames, IA  50011,
$^m$University of Iowa, Iowa City, IA  52242,
$^n$Kinki University, Higashi-Osaka City, Japan 577-8502,
$^o$Kansas State University, Manhattan, KS 66506,
$^p$University of Manchester, Manchester M13 9PL, England,
$^q$Queen Mary, University of London, London, E1 4NS, England,
$^r$University of Melbourne, Victoria 3010, Australia,
$^s$Muons, Inc., Batavia, IL 60510,
$^t$Nagasaki Institute of Applied Science, Nagasaki, Japan, 
$^u$National Research Nuclear University, Moscow, Russia,
$^v$University of Notre Dame, Notre Dame, IN 46556,
$^w$Universidad de Oviedo, E-33007 Oviedo, Spain, 
$^x$Texas Tech University, Lubbock, TX  79609, 
$^y$Universidad Tecnica Federico Santa Maria, 110v Valparaiso, Chile,
$^z$Yarmouk University, Irbid 211-63, Jordan,
$^{hh}$On leave from J.~Stefan Institute, Ljubljana, Slovenia, 
}}
\noaffiliation

\begin{abstract}
 We present a search for a new particle $T'$ decaying to top quark via
 $T^{\prime} \rightarrow t + X$, where $X$ is an invisible particle. In a
 data sample with 4.8 fb$^{-1}$ of integrated luminosity collected by
 the CDF II detector at Fermilab in $p\bar{p}$ collisions with
 $\sqrt{s}=1.96$ TeV, we search for pair production of $T'$ in the lepton+jets channel, $p\bar{p}
 \rightarrow t\bar{t} +X+X \rightarrow \ell\nu b q q' b + X +X$.  We
 interpret our results primarily in terms of a model where $T'$ are exotic fourth generation quarks and $X$ are dark matter particles. Current direct and indirect bounds on such exotic quarks restrict their masses to be between 300 and 600 GeV$/c^2$, the dark matter particle mass being anywhere below $m_{T^{\prime}}$.  The data are consistent with standard model expectations, and we set 95\% confidence level limits on the
generic production of $T'\bar{T'} \rightarrow t\bar{t}+X+X$. We apply these limits to the dark matter model and exclude the
fourth generation exotic quarks $T'$ at 95\% confidence level up to
$m_{T'}=360$ GeV$/c^{2}$ for $m_X\le 100$ GeV$/c^2$.
\end{abstract}



\maketitle

\date{\today}

Despite an intensive program of research~\cite{dmsearch}, the precise nature of dark matter
remains elusive, though it is clear that it must be long-lived on
cosmological time scales. Such a long lifetime could be due to a
conserved charge under an unbroken symmetry.  However, none of the the unbroken
symmetries of the standard model (SM) suffice to provide such a charge, so
it follows that dark matter must be charged under a new, unbroken
symmetry.  The prospects of creating dark matter at particle colliders
are excellent, but only if the dark matter particles $X$ couple to
standard model particles directly or indirectly. One potential
mechanism is via a connector particle $Y$, which carries SM charges so that it can be produced at particle colliders as well as carrying the new dark charge, so that it can decay to the dark matter particle, $Y\rightarrow f+X$, where $f$ is a SM particle.  One compelling recent model~\cite{feng} uses an exotic fourth generation up-type quark $T'$ as the connector particle, which decays to a top quark and dark matter, $T'\rightarrow t+ X$. Current direct and indirect bounds on such exotic quarks restrict their masses to be between 300 and 600 GeV~\cite{feng}.  

The pair production of such exotic quarks and their subsequent decay
to top quarks and dark matter has a collider signal comprising of top
quark pairs($t\bar{t}$) and missing transverse momentum($\slashed{E}_{t}$) due to the invisible dark matter particles. These types of signals, in general, are of great interest as they appear in numerous new physics scenarious including many dark matter motivated models, little Higgs models with $T$-parity conservation~\cite{lhiggs} and models in which baryon and lepton numbers are gauge symmetries~\cite{blmodels}. Supersymmetry, which includes a natural dark matter candidate and provides a framework for unification of the forces, also predicts a $t\bar{t}+\slashed{E}_{t}$ signal from the decay of a supersymmetric top 
$\tilde{t}$ quark to a top quark and the lightest
supersymmetric particle~\cite{susy}, $\tilde{t}\rightarrow
t+\chi^0$. There are currently no experimental bounds on a new heavy
particle $Y$ decaying via $Y \rightarrow t+X$.

This Letter reports a search for such a generic signal $t\bar{t}+\slashed{E}_{t}$ via the pair production of a heavy new
particle $T'$ with prompt decay $T' \rightarrow t+X$. We consider the mode $p\bar{p}
\rightarrow t\bar{t} +X+X \rightarrow WbWb + X +X$ in which one $W$
decays leptonically (including $\tau$ decays to $e$ or $\mu$) and one
decays hadronically to $qq'$, this decay mode allows for large branching ratios while constraining SM backgrounds. Such a signal is similar to top quark
pair production and decay, but with additional missing transverse energy due to the invisible particles.

Events were recorded by CDF II~\cite{cdf}, a general purpose detector
designed to study collisions at the Fermilab Tevatron
$p\overline{p}$ collider at $\sqrt{s}=1.96$ TeV. A charged-particle tracking system immersed
in a 1.4~T magnetic field consists of a silicon microstrip tracker and
a drift chamber. Electromagnetic and hadronic calorimeters surrounding
the tracking system measure particle energies and drift chambers
located outside the calorimeters detect muons.  We use a data sample
corresponding to an integrated luminosity of 4.8$\pm 0.3$\invfb.

The data acquisition system is triggered by $e$ or $\mu$ candidates~\cite{leptonid}
with transverse momentum $p_T$\cite{coordinates}  greater than $
18$\gevc.  Electrons and muons are reconstructed offline and are selected if they
have a pseudorapidity $\eta$\cite{coordinates} magnitude less than
1.1, $p_T \ge 20$ GeV$/c$ and satisfy the standard identification and
isolation requirements~\cite{leptonid}.  Jets are reconstructed in the
calorimeter using the {\sc jetclu}~\cite{jetclu}  algorithm with a
clustering radius of 0.4 in azimuth-pseudorapidity space and corrected using standard techniques~\cite{jes}. Jets are
selected if they have $p_T \ge 15$ GeV$/c$ and $|\eta| < 2.4$. Missing transverse momentum~\cite{metdef}
is reconstructed using fully corrected
calorimeter and muon information~\cite{leptonid}.

Production of $T'$ pairs and their subsequent decays to top quark pairs and
two dark matter particles would appear as events with a charged lepton
and missing transverse momentum from one leptonically decaying $W$ and
the two dark matter particles, and four jets from the two $b$ quarks
and the hadronic decay of the second $W$ boson. We select events with at least one electron or
muon, at least four jets, and large
missing transverse momentum. The missing transverse energy in a
signal event depends on the masses $m_{T'}$ and $ m_{X}$, for each
pair of signal masses we optimize for the  minimum amount of missing
transverse energy required(ranging from 100 GeV$/c$ to 160 GeV$/c$).

We model the production and decay of $T'$ pairs with {\sc
  madgraph}~\cite{madgraph}. Additional radiation, hadronization and
showering are described by {\sc pythia}~\cite{pythia}. The detector
response for all simulated samples is modeled by the official CDF detector simulation.

The dominant SM background is top-quark pair production. We model this background using
{\sc pythia} $t\bar{t}$ production with $m_t = 172.5$ GeV/$c^2$. We
normalize the $t\bar{t}$ background to the NLO cross
section~\cite{nloa}, and confirm that it is well modelled by examining $t\bar{t}$-dominated regions in the data.

 The second dominant SM background process is the associated production 
of $W$ boson and jets. Samples of simulated $W$+jets events with light- and heavy-flavor 
jets are generated using the {\sc alpgen}~\cite{alpgen} program, interfaced with parton-shower model 
from {\sc pythia}.  The $W+$jets samples are normalized to the measured $W$ cross section, 
with an additional multiplicative factor for the relative contribution of heavy- 
and light-flavor jets, the standard technique in measuring the
top-quark pair production cross section~\cite{secvtx}. Multi-jet
background, in which a jet is
misreconstructed as a lepton, is modeled using a jet-triggered sample normalized
in a background-dominated region at low missing transverse momentum.
The remaining backgrounds, single top and diboson production, are modeled
using {\sc pythia} and normalized to next-to-leading order
cross sections~\cite{ewknlo}. 

\begin{figure}[!h]
\begin{center}
\includegraphics[width=0.8\linewidth]{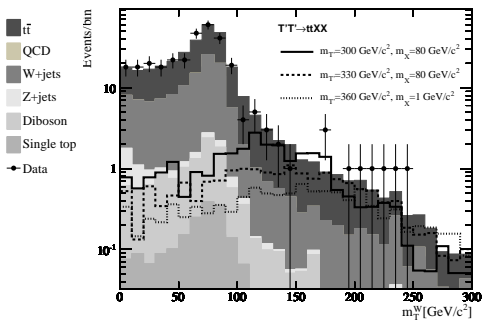}\\
\caption{ Reconstructed transverse mass of the $W$, $m_T^{W}$, for
  the standard model backgrounds, the observed data, and for three
  choices of ($m_{T'},m_{X}$).}
\label{fig:signals}
\end{center}
\end{figure}

We differentiate the signal events from these backgrounds by comparing
the reconstructed transverse mass of the leptonically decaying $W$
candidate,

\[ m_T^W\equiv
m_{T}(p_{T}^{\ell},\slashed{p}_{T})=\sqrt{2|p_{T}^{\ell}||\slashed{p}_{T}|(1-
 cos(\Delta\phi(p_{T}^{\ell},\slashed{p}_{T})))}. \]

\noindent
where $p_T^\ell$ is the transverse momentum of the lepton and
$\slashed{p}_{T}$ is the missing transverse momentum.  In background
events, the $\slashed{p}_{T}$ comes primarily from the neutrino in
$W\rightarrow \ell\nu$ decay, and $m_T^W$ will show a strong peak at
the $W$-boson mass. The signal event, $T^{\prime} \rightarrow t + X$, has additional missing transverse momentum due to the invisible particles X and thus does not reconstruct the W-mass in $m_T^W$. Figure~\ref{fig:signals} show the $m_T^{W}$ distributions of the backgrounds versus the signals.

We consider several sources of systematic uncertainty on both the background
rates and distributions, as well as on the expectations for the
signal. Each affects the expected sensitivity to new physics expressed
as an expected cross section upper limit in the no-signal
assumption. The dominant systematic uncertainties are the jet energy scale~\cite{jes},
contributions from additional interactions, and descriptions of
initial and final state radiation~\cite{rad}.  In each case, we treat
the unknown underlying quantity as a nuisance parameter and measure the distortion of the
 $m_T^{W}$ spectrum for positive and negative fluctuations. As mentioned before we optimize the minimum missing transverse energy required for each signal point, Table~\ref{Table2} compares the number of events expected with uncertainties for backgrounds and signals to data for two example missing transverse energy cuts.

\begin{center}
\begin{table}[!h]
\renewcommand{\arraystretch}{1.2}
\caption{Number of events for example signal points compared to backgrounds and data for two $\slashed{E}_{T}$ cuts after initial selection is made. }
\begin{ruledtabular}
\begin{tabular}{c c c} 
Cut:&$\slashed{E}_{T}\geq 100$ GeV$/c$&$\slashed{E}_{T}\geq 150$ GeV$/c$\\\hline
$T^{\prime}T^{\prime}\rightarrow ttXX$ [GeV$/c^{2}$]&  \\ 
$m_{T^{\prime}},m_{X}=300,90$&  $22.9^{+5.8}_{-4.7}$&  $4.1^{+2.4}_{-2.1}$   \\ 
$m_{T^{\prime}},m_{X}=310,80$&  $22.6^{+4.9}_{-5.1}$&  $6.4^{+2.3}_{-2.6}$  \\ 
$m_{T^{\prime}},m_{X}=330,70$&  $17.6^{+3.7}_{-3.6}$&  $7.3^{+2.5}_{-2.4}$  \\  
$m_{T^{\prime}},m_{X}=350,1$&   $13.1^{+2.7}_{-2.8}$&  $6.7^{+2.0}_{-1.9}$\\ \hline
t\={t}&   $189^{+54}_{-50}$&   $26.3^{+11.6}_{-9.8}$ \\ 
W+jets& $105^{+31}_{-14}$&   $16.6^{+4.5}_{-2.1}$ \\ 
Single top & $1.86\pm 0.2$&   $0.18\pm 0.02$  \\
Diboson &  $9.69\pm 0.1$&   $1.53\pm 0.1$  \\ 
Z+jets &   $4.00\pm 0.4$&   $0.46\pm 0.05$   \\ 
QCD  &   $0.04\pm 0.01$&   $0.04\pm 0.01$  \\ \hline
Total Background &   $310^{+80}_{-64}$&   $45^{+14}_{-11}$ \\ \hline\hline
Data  &   309 &   42 \\ 
\end{tabular}
\end{ruledtabular}
\label{Table2}
\end{table}
\end{center}

\begin{figure}[!htb]
\begin{center}
\includegraphics[width=1.0\linewidth]{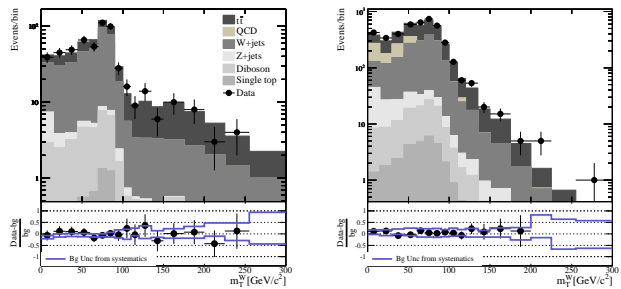}\\
\caption{Reconstructed transverse mass of the $W$, $m_T^{W}$, in
  signal-depleted control regions. Left, events with at least four
  jets and small missing transverse momentum ($<100$ GeV$/c$). Right,
  events with exactly three jets and large missing transverse momentum ($>100$ GeV$/c$).}
\label{fig:mtw}
\end{center}
\end{figure}

We validate our modeling of the SM backgrounds in two
background-dominated control regions. We validate our modeling of the
large $m_T^{W}$ region in events with high missing transverse energy
and exactly three jets, and validate our modeling of four-jet events
in events with small missing transverse energy ($<100$ GeV$/c$). Figure.~\ref{fig:mtw} shows good agreement of our background modeling with data in the control regions.

\begin{figure}[!h]
\begin{center}
\includegraphics[width=0.8\linewidth]{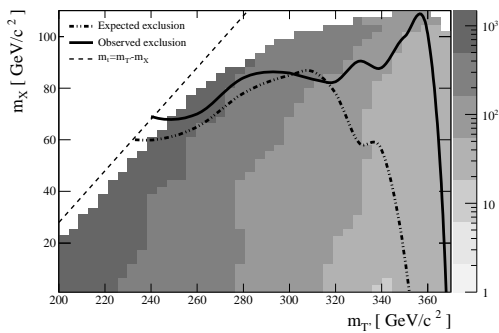}\\
\caption{Observed versus expected exclusion in ($m_{T^{\prime}},m_{X}$) along with the cross section upper limits.}
\label{fig:sigmadataexclusion}
\end{center}
\end{figure}

There is no evidence for the presence of $T^{\prime} \rightarrow t + X$ events in the data. We calculate 95\% C.L. upper limits on the $T^{\prime} \rightarrow t + X$ cross section, by performing a binned maximum-likelihood fit in the $m_T^{W}$ variable, allowing for systematic and statistical fluctuations via template morphing~\cite{morph}. We use the likelihood-ratio ordering
prescription~\cite{feldcous} to construct classical confidence
intervals in the theoretical cross section by generating ensembles of
simulated experiments that describe expected fluctuations of
statistical and systematic uncertainties on both signal and
backgrounds. The observed limits are consistent with expectation in
the background-only hypothesis, for a few example signal mass points we tabulate the expected and observed limits(see Table~\ref{Table1}). We convert the observed upper limits on the pair-production cross sections to an exclusion curve in mass parameter space for the dark matter model
involving fourth generation quarks, see Fig.~\ref{fig:sigmadataexclusion}.

\begin{table}[h!]
\caption{Expected 95\% CL upper limit on $T'\bar{T'}$ production cross-section,
    $\sigma_{exp}$, the range of expected limits which includes 68\%
    of pseudoexperiments, and the 
                              observed limit, $\sigma_{obs}$, for representative signal points in ($m_{T^{\prime}},mX$).  }  
\begin{minipage}[h]{0.45\textwidth}
\centering
\begin{ruledtabular}
\begin{tabular}{p {1.5 cm}p {1.5 cm} c  cp {1.5 cm}}
$m_{T^{\prime}}$,$m_{X}$ (GeV$/c^{2}$)& $\sigma_{exp}$ [pb] & $+34\%$ & $-34\%$ & $\sigma_{obs}$ [pb].\\ \hline 
200,1  & 1.31 & 1.86 & 0.83 & 1.21\\ \hline  
220,40  & 1.40 & 2.17 & 0.93 &  1.20\\ \hline 
260,1  & 0.23 & 0.40 & 0.14 & 0.20\\ \hline 
280,1  & 0.16 & 0.27 & 0.09 &  0.15\\ \hline 
280,20  & 0.18 & 0.29 & 0.11 &  0.17\\ \hline 
280,40 & 0.17 & 0.27 & 0.11 &  0.12\\ 
\end{tabular}
\end{ruledtabular}
\end{minipage}
\hspace{0.1cm}
\begin{minipage}[h]{0.45\textwidth}
\centering
\begin{ruledtabular}
\begin{tabular}{p {1.5 cm}p {1.5 cm} c  cp {1.5 cm}}
$m_{T^{\prime}}$,$m_{X}$ (GeV$/c^{2}$)& $\sigma_{exp}$ [pb] & $+34\%$ & $-34\%$  &$\sigma_{obs}$ [pb].\\ \hline 
300,100  & 0.34 & 0.51 & 0.24 & 0.39\\ \hline 
310,90 & 0.19 & 0.29 & 0.11 &  0.21\\ \hline 
320,80  & 0.15 & 0.24 & 0.08 &  0.12\\ \hline
350,50 & 0.07 & 0.10 & 0.04 &  0.02\\ \hline 
360,110 & 0.09 & 0.19 & 0.05 &  0.09\\ \hline 
370,1  & 0.06 & 0.10 & 0.04 &  0.05\\ 
\end{tabular}
\end{ruledtabular}
\label{Table1}
\end{minipage}
\end{table}

In conclusion, we have searched for new physics
particles $T'$ 
decaying to top quarks with invisible particles $X$ with a detector
signature of  $t\bar{t}+\slashed{E}_{t}$. We calculate upper limits on
the cross section of such events and exclude a dark matter model
involving exotic fourth generation quark up to $m_{T'}=360$
GeV$/c^{2}$. Our cross section limits on the generic decay, $T^{\prime} \rightarrow t +
X$, may be applied to the many other models that predict the
production of a heavy particle $T'$ decaying to top quarks and
invisible particles $X$,  such
as the supersymmetric process
$\tilde{t}\rightarrow t+\chi^0$. A similar search performed at the LHC, given its higher energy regime, would be able to provide limits on such a supersymmetric decay.

We thank the Fermilab staff and the technical staffs of the participating institutions for their vital contributions. This work was supported by the U.S. Department of Energy and National Science Foundation; the Italian Istituto Nazionale di Fisica Nucleare; the Ministry of Education, Culture, Sports, Science and Technology of Japan; the Natural Sciences and Engineering Research Council of Canada; the National Science Council of the Republic of China; the Swiss National Science Foundation; the A.P. Sloan Foundation; the Bundesministerium f\"ur Bildung und Forschung, Germany; the World Class University Program, the National Research Foundation of Korea; the Science and Technology Facilities Council and the Royal Society, UK; the Institut National de Physique Nucleaire et Physique des Particules/CNRS; the Russian Foundation for Basic Research; the Ministerio de Ciencia e Innovaci\'{o}n, and Programa Consolider-Ingenio 2010, Spain; the Slovak R\&D Agency; and the Academy of Finland. 


\begin{thebibliography}{99}
\bibitem{coordinates}
CDF uses a cylindrical coordinate system with
  the $z$ axis along the proton beam axis. Pseudorapidity is $\eta
  \equiv -\ln(\tan (\theta/2))$, where $\theta$ is the polar angle
  relative to the proton beam direction, and $\phi$ is the azimuthal angle  while $p_T = |p|\ \sin\theta$, $E_T = E\ \sin\theta$.
\bibitem{metdef} 
Missing transverse momentum, $\missET$, is defined as the magnitude of the vector $ -\sum_i E_T^i \vec n_i$ where $E_T^i$ are the magnitudes of transverse energy contained in each calorimeter tower $i$, and $\vec n_i$ is the unit vector from the interaction vertex to the tower in the transverse ($x,y$) plane.
\end{thebibliography}

\end{document}